# Ontology Based Pivoted normalization using Vector – Based Approach for information Retrieval

Vishal Jain[1] and Dr. Mayank Singh[2]

[1]Research Scholar, Computer Science and Engineering Department, Lingaya's University, Faridabad
[2]Associate Professor, Computer Science and Engineering Department, Lingaya's University, Faridabad
[1]vishaljain83@ymail.com, [2]mayanksingh2005@gmail.com

## ABSTRACT

*An ample amount of documents present on web puts the users in state of dilemma. Users get confused about relevance of documents. Relevance means how closely the given query matches large number of documents. Many information extraction techniques are used for extracting documents but they all are in vain. The paper deals with the problem of classification, analyzing and extraction of web documents by using one of information extraction methods called Ontology Based Web Content Mining Methodology. We have evaluated proposed methodology in two specific domains- weather domain (web pages containing information about weather forecasting and analysis) and Google TM collection (web pages containing news).*

*The proposed methodology is procedural i.e. it follows finite number of steps that extracts relevant documents according to user's query. It is based on principles of Data Mining for analyzing web data. Data Mining first adapts integration of data to generate warehouse. Then, it extracts useful information with the help of algorithm. The task of representing extracted documents is done by using Vector Based Statistical Approach that represents each document in set of Terms.*

## 1. INTRODUCTION

Data Mining is called as Knowledge Discovery in Databases (KDD) [1]. It is multi-level field i.e. it includes different areas like Database Systems, Information Retrieval (IR), Machine Learning etc. Prediction and Description are considered as two goals of Data Mining where Prediction involves use of some variables or records in database to predict future values of other variables while Description finds useful patterns describing the given data.

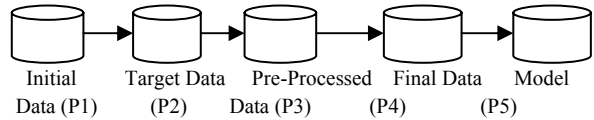

Initial Data (P1)   Target Data (P2)   Pre-Processed Data (P3)   Final Data (P4)   Model (P5)

**Figure 1: KDD Process [2]**

Building Ontology needs attention of domain expert that represents concepts and relations between them for a given domain. There are many algorithms used for extracting and discovering knowledge from structured data like Naïve Bayes, K-Means etc. The proposed methodology builds ontology for a given domain by using phases of data mining like Data preparation, Data Mapping, extracting knowledge from mapped data etc. Then, classification algorithm is used for writing generated ontology expressed in OWL and XML languages.

There are various uses of Ontology:

- Used for knowledge sharing and reuse.
- Can improve understanding between concepts.
- It is useful in Semantic Web that is information in machine form.
- Some search engines use ontology for finding relevant pages related to given query.

The paper is divided into following sections: Section 2 gives information about following concepts:

- Domain Ontology
- Stages of Ontology Based Web Content Mining Methodology
- Increasing accuracy of classification of web documents using WORDnet.

Section 3 describes representation schemes of documents extracted during Ontology Based Phrase Extractor by using Statistical Vector Based Approach. Section 4 concludes about given paper.

## 2. CLASSIFYING AND ANALYSING WEB DOCUMENTS

Many information extraction methods and techniques were used but they all are in vain. So we need more intelligent system to gather useful information from huge amount of data.

*Problem:* - To find meaningful and informative documents with help of Data Mining algorithms and then interpreting mining results in expressive way.

*Solution*: - Ontology Based Web Content Mining Methodology

*Approach involved*: - The proposed methodology uses concept of *Domain Ontology* [3]. Domain Ontology organizes concepts, relations and instances into given domain. This approach is used because it resolves synonyms and reducing confusion among agents

### 2.1 Ontology Web Based Content Mining

Ontology Based Web Content Mining represents conceptual information about given domain. It shows document representation, extraction of relevant information from text documents and creates classification models. This methodology is followed that uses the ideas and principles of Data Mining to analyze web data.

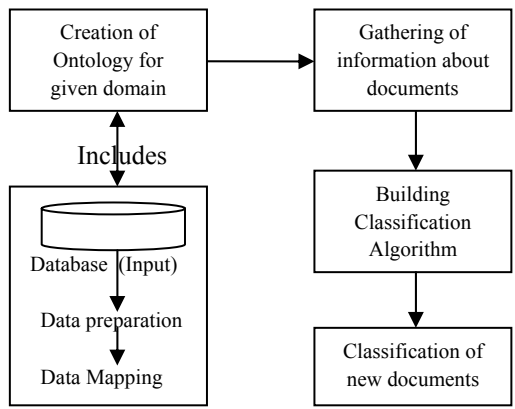

**Figure 2: Stages of Ontology Web Based Content Mining**

### 2.2 Building Ontology for given domain

*Importance*: - Since traditionally domain experts were not so intelligent that they could represent complete knowledge related to query. So, there is need for updating knowledge frequently. It leads to building of ontology.

It takes data from given database located at back end server. Data Preparation is included in order to completely understand the meaning of data and lists all tables and attributes that are present in database. Data Mapping states that data is to be represented according to some algorithm. Mapper is used for data mapping. Mapper converts Input data into normal format so that it satisfies user's requirements in Building Classification algorithm phase.

### 2.3 Gathering of Information about Documents

It involves use of Ontology Based Phrase Extractor. Its specification is as follows:

- Input = Web documents + Domain Ontology and User Abstraction level (K)
- Output = Documents associated with vector terms ($t_i$) and weights ($w_i$).

*Process*: - Extractor prepares XML file containing instances of ontology with their relationships in hierarchy level. In WORDnet, phrase collection means relevant phrases with their associated concepts of ontology.

To extract concepts, we use disambigutive function dis (t) that shows semantically concept for terms ($t_i$) based on given topic. Phrase Extractor as name suggests scans the phrases and as it finds some relevant matter, it refers to related concepts. Each web document is represented as vector of < term $t_i$>, <weight $w_i$> pairs which is extracted from Phrase extractor module.

### 2.4 Classification Algorithm

This ontology building algorithm [4] is written on basis of decision tree as follows:

| INPUT | OUTPUT |
|---|---|
| (i) Decision Tree<br>(ii) Distinct Nodes<br>(iii) Distinct Tree branches<br>(iv) Target Attribute<br>(v) GetBranches ()- It is function to get all branches having given node<br>(vi) GetLeafBranch ()- A function to get branch of leaf node<br>(vii) GetClass ()- To get class that shows tree branch<br>(viii) CreateIndividual ()- To create instance of leaf node | Ontology |

Algorithm is as follows:
Begin
For each node N of decision nodes
Class C = new (owl: Class)
C.Id = N.name
DatatypeProperty DP = new (owl: DatatypeProperty)
DP.Id = N.name + "Value"
Dp.AddDomain (C);    // It will add property of Class c in its child node//
For each branch B of GetBranches (N)
Dp.AddDomain (B.GetClass (C))
End for
End

*Working*: - The algorithm generates ontology written in OWL language. It creates class for each distinct node and assigns them with unique ID and name. Each child node of specific parent node is assigned name and value by acquiring property of parent node using DatatypeProperty DP class.

All the branches including specific node are returned using GetBranches () function. For generating ontology, traverse decision tree from root node till we get all unique nodes.

## 2.4 Increasing accuracy of classification of web documents using WORDnet

*About WORDnet*: - WORDnet was invented by A. Miller [5]. It is a large lexical database of English. Nouns, Verbs, Adverbs, Adjectives combine into sets of synonyms. Each of synonyms represents concept and solves queries through search and lexical results. In this paper, we have used WORDnet version 3.1. It contains 155287 synonyms. WORDnet is one of best example of ontology used in experiments

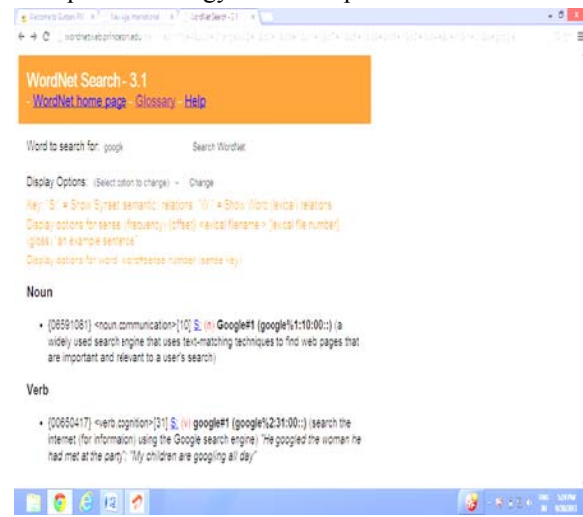

### 2.4.1 Experiment

This experiment is done to improve accuracy in classification of web documents using Domain Ontology. We tested system in two domains- weather forecasting and GoogleTM. *There is specific abstraction level (k) related to each domain*. Our system is tested for both domains before and after abstraction. We have used WORDnet for experimental evaluations.

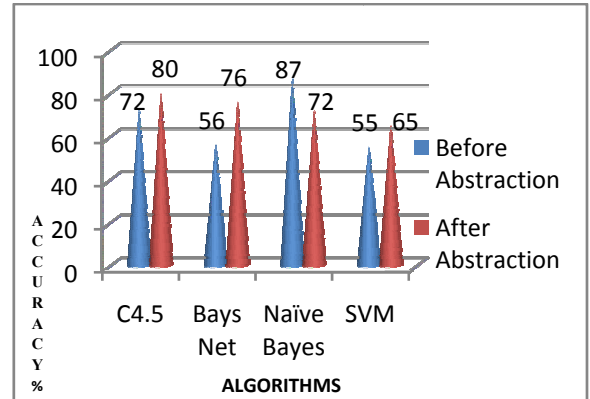

*Results*- The above experiment shows that classification Data Mining algorithms like C4.5, Bayes Net and SVM are improved by using WORDnet. We can classify simple words and expressions of different datasets. *Naïve Bayes* is algorithm that shows accurate result before abstraction and shows less accurate result after abstraction

## 3. REPRESENTATION OF EXTRACTED DOCUMENTS USING VECTOR BASED APPROACH

Vector based approach is one of statistical approaches. Such approaches break documents or given query into TERMS. Terms are words that occur in given query and are extracted automatically form documents. These words are counted and measured statistically.

Our aim is to remove different forms of same word. E.g. play is word entered by user. It has various forms like played, playing etc. User can specify only one form of above word in searching query.

### 3.1 Relation between Term Vectors in Document Space

The terms can be phrases, n-grams etc. We will represent document as set of terms. Take ORing of these terms. We get set of terms that represents entire document called as Space.

$T_1$ OR $T_2$ OR $T_3$ OR …..$T_n$ = Space

A document consisting set of terms (space) is called Document Space. Numeric weights are assigned to each term in document that estimates effectiveness of document comparing them with other documents. Each term has different weight in same document. The weights assigned to each term in document $D_i$ are expressed as coordinates of document i.e. $D_i$ (x,y). So, it is called as Vector from origin to point defined by weight of terms.

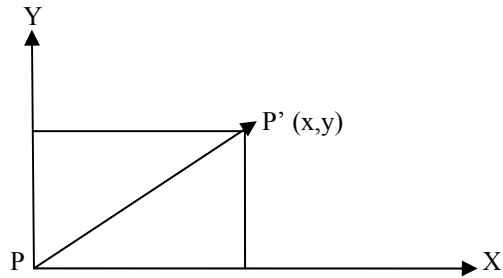

Term Space: - Each document is represented as dimension. It has some coordinates (weights). Each point is considered as vector. If term is not found in document, then it is assigned as zero weight.

*Representation of terms in matrix form*
Combination of document space and term space is represented by Document - by Term Matrix. In this, each row is document $D_i$ (in term space) and each column is term (in document space). [ ] Di x Term

*Representation of Query in document space*
A query entered by user is a set of terms having same weights assigned to it. Query may be in natural language also. In this, it is processed like document which includes removal of redundant words.
If query contain terms that are not in document, then it represent dimensions in document space.

### 3.2 Assignment of weights to terms
Weight of terms means importance of term i.e. how relevant it is. Weights are assigned by special scheme called as Term Frequency * Inverse Document Frequency (tf * idf)
Term Frequency (tf): - It defines number of terms occurred in document. So, it varies from one document to another.
Inverse Document Frequency (idf): - It means how many times the given term is distributed in document. It gives probability of terms occurred in document.
**idf= ln N/n**
**where N= number of documents**
**n = number of relevant documents**

*Inference*: - If all documents are relevant, then idf is zero. We can say that for distinguishing relevant and non relevant documents, the terms in document must be different from given topic so that they can be used for comparing with other documents.

*Why idf is multiplied by tf?*
It is done so that good descriptor terms have more importance than bad terms. Good terms are those that occur in small number of documents while Bad terms are those that occur in large number of documents.

### 3.3 Normalization of Term Vectors
Weights are normalized according to variable document size. Here we will describe Normalization of term frequency (tf).
In this, tf is divided by maximum term frequency $tf_{max}$ i.e. $tf/tf_{max}$. It is defined as frequency of term that occurs mostly in documents. So, we generate factor that lies between 0 and 1. This kind of normalization is called as Maximum Normalization. It is given as:
**$[0.5 + (0.5 * tf/tf_{max})]$** where tf varies from 0.5 to 1.
*Effect of tf*: - The importance of term in given document depends on its frequency of occurrence as compared to other terms in same document. Terms are variables. They can change anytime.
*Drawback*: - Since, normalization factor depends only on frequency of documents. The problem is that terms having higher weights can replace terms with lower weights. E.g. A document is about computer design. It includes various components, hardware software. Let us consider hardware is highly weighted term that occurs six times in document. It will occur most because it is used in building computer. Then, the frequency of this term will replace all other terms by factor of 3.
*Solution*: - Logarithmic Term Frequency
In this, we take natural log plus constant i.e. log (tf) +1. Its normalization factor does not depend on maximum term frequency ($tf_{max}$). It reduces the effect of term with high frequency such that two terms $Tf_1$ & $Tf_2$ >0 then
**$[Log (tf_2) +1 / log (tf_1) +1] < tf_2 / tf_1$**.

### 3.3.1 Normalization by Vector Length
In this, every component of vector is calculated. Each component is divided by Euclidian length of vector.

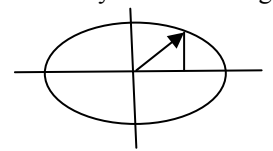

R = x i + y j
Euclidian Length of R = $\sqrt{x^2 + y^2}$
Cosine Normalization = $x / \sqrt{x^2 + y^2}, y / \sqrt{x^2 + y^2}$
= $x / \sqrt{r^2\cos^2 + r^2\sin^2}$

It is called cosine normalization because normalized vector $\sqrt{\cos^2 + \sin^2}$ has length = 1. It is written as n^=1. Cosine normalization reduces the effect of single term with high frequency by combining it with other low weighted terms. Since vector length is function of all vector components i.e. tdf *idf weights. So, weight of high frequency term is reduced by idf factor.

Cosine normalization takes into account the weights of all terms in a given document. It is done for short documents rather than longer documents because short documents are about single topic relevant to given query.

For every document or query, there is a stage when all the terms that are retrieved are also relevant. It is indicated by probability of relevance = Probability of retrieval. It has led to development of *Correction Factor*. It is factor that maps old normalization function (Cosine normalization) into new function.

*Concept of Pivot Normalization*

This correction factor rotates the old normalized function clockwise around crossover point (point where probability of relevance = Probability of retrieval) so that normalization values below that point are greater and values above it are lesser. The crossover point is called PIVOT. It is called as Pivot Normalization.

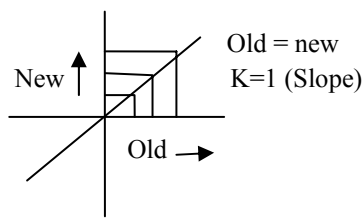

Pivot normalization focuses on correcting the document normalization. Before pivoted normalization, old normalization = new normalization. When it is rotated clockwise around pivot, then **new normalization = Slope * old normalization + constant**. Here slope should be less than 1. Putting arbitrary value of pivot for both old and new normalizations, we get final result i.e.

**[Pivoted normalization = Slope * old normalization + (1 – slope) * Pivot].**

Use of pivoted normalization increases probability of retrieving longer documents although longer documents have both relevant s well as non relevant terms.

## 4. CONCLUSION

The paper presents Ontology Web Based Content Mining methodology that helps in classification, identification and extraction of large number of documents present on web. It follows certain number of steps for generating ontology. We have conducted an experiment using WORDnet. The main credit of this work goes to domain ontology in representing documents. Use of WORDnet leads to improvement in classification of web documents with the help of synonyms as it has large collection of similar words related to particular search.

Ontology Phrase extractor produces web documents that consists of multiple pages with multiple categories. Each web document is represented as vector of < term $t_i$>, <weight $w_i$> pairs. They are represented using one of statistical Information retrieval (IR) approaches known as Vector- based Approach. The paper also represents terms in document space and normalized them using concept of Pivoted normalization.

**ACKNOWLEDGEMENT**

I, Vishal Jain would like to give my sincere thanks to Prof. M. N. Hoda, Director, Bharati Vidyapeeth's Institute of Computer Applications and Management (BVICAM), New Delhi for giving me opportunity to do Ph.D from Lingaya's University, Faridabad.

## About the Authors

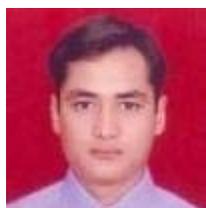
**Vishal Jain** has completed his M.Tech (CSE) from USIT, Guru Gobind Singh Indraprastha University, Delhi and doing PhD from Computer Science and Engineering Department, Lingaya's University, Faridabad. Presently he is working as Assistant Professor in Bharati Vidyapeeth's Institute of Computer Applications and Management, (BVICAM), New Delhi. His research area includes Web Technology, Semantic Web and Information Retrieval. He is also associated with CSI, ISTE.

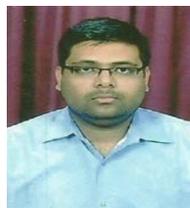
**Dr. Mayank Singh** has completed his M. E in software engineering from Thapar University and PhD from Uttarakhand Technical University. His Research area includes Software Engineering, Software Testing, Wireless Sensor Networks and Data Mining. Presently He is working as Associate Professor in Krishna Engineering College, Ghaziabad. He is associated with CSI, IE (I), IEEE Computer Society India and ACM.